\documentclass[journal=jctcce,manuscript=article]{achemso}

\usepackage[version=3]{mhchem} %
\usepackage{enumitem}
\usepackage{amsmath}
\usepackage{amssymb}
\usepackage{amsfonts}
\usepackage{booktabs}
\usepackage{xurl}
\usepackage[table]{xcolor}

\author{Soojung Yang}
\affiliation{Computational and Systems Biology Program, Massachusetts Institute of Technology, Cambridge, MA 02139, USA}
\altaffiliation{Contributed equally to this work}
\author{Juno Nam}
\affiliation{Department of Materials Science and Engineering, Massachusetts Institute of Technology, Cambridge, Massachusetts 02139, United States}
\altaffiliation{Contributed equally to this work}
\author{Johannes C. B. Dietschreit}
\affiliation{Department of Materials Science and Engineering, Massachusetts Institute of Technology, Cambridge, Massachusetts 02139, United States}
\alsoaffiliation{Institute of Theoretical Chemistry, Faculty of Chemistry, University of Vienna, 1090 Vienna, Austria}
\author{Rafael G{\'o}mez-Bombarelli}
\affiliation{Department of Materials Science and Engineering, Massachusetts Institute of Technology, Cambridge, Massachusetts 02139, United States}
\email{rafagb@mit.edu}

\title{Learning Collective Variables with Synthetic Data Augmentation through Physics-Inspired Geodesic Interpolation}
\abbreviations{}
\keywords{Protein folding, Molecular dynamics, Data augmentation, Geodesic interpolation, Collective variables, Enhanced sampling}

\begin{document}

\begin{tocentry}
\includegraphics{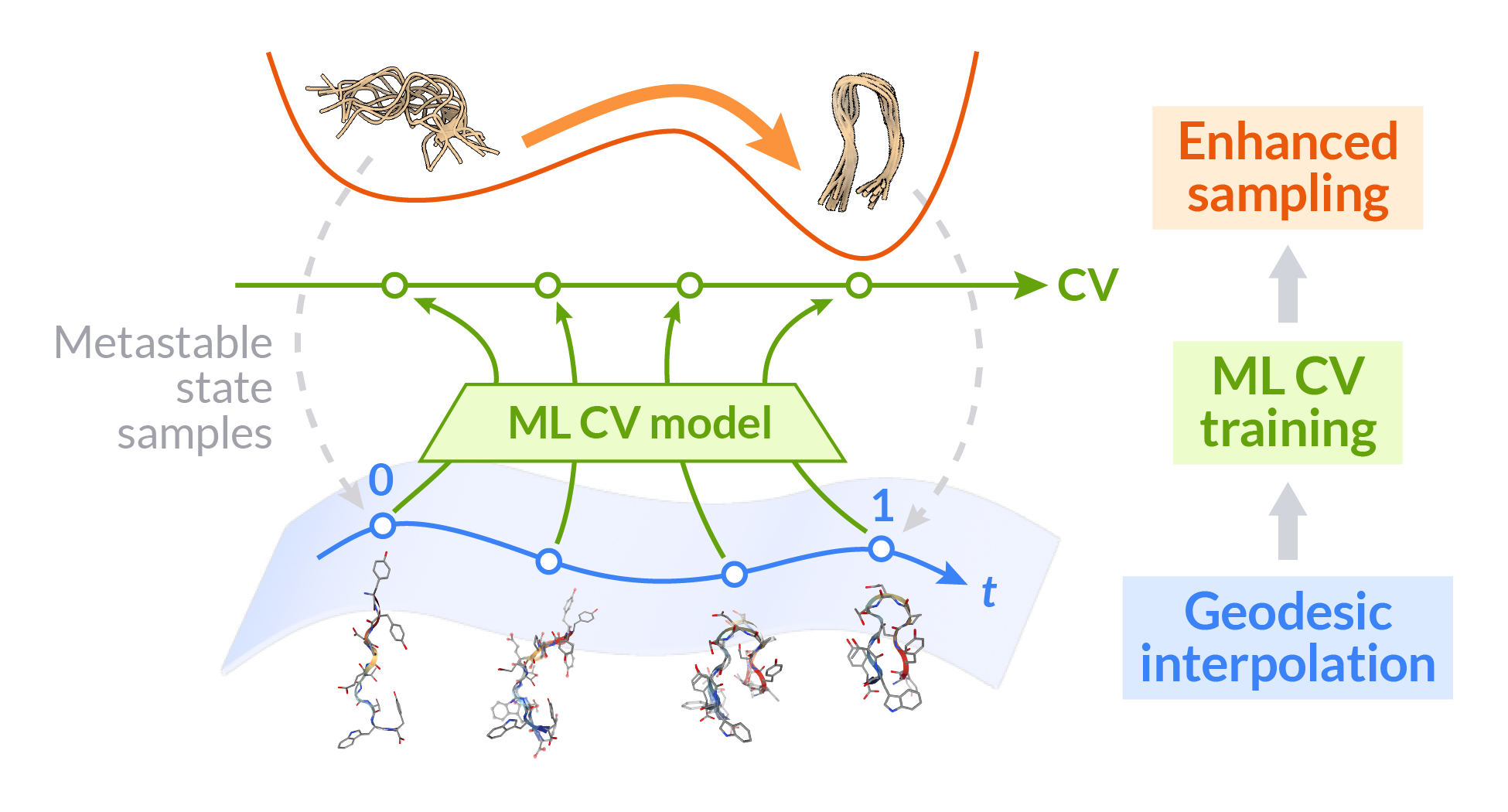}
\end{tocentry}

\begin{abstract}
In molecular dynamics simulations, rare events, such as protein folding, are typically studied using enhanced sampling techniques, most of which are based on the definition of a collective variable (CV) along which acceleration occurs.
Obtaining an expressive CV is crucial, but often hindered by the lack of information about the particular event, e.g., the transition from unfolded to folded conformation.
We propose a simulation-free data augmentation strategy using physics-inspired metrics to generate geodesic interpolations resembling protein folding transitions, thereby improving sampling efficiency without true transition state samples. 
This new data can be used to improve the accuracy of classifier-based methods.
Alternatively, a regression-based learning scheme for CV models can be adopted by leveraging the interpolation progress parameter. 
\end{abstract}

\section{Introduction}
\label{Introduction}

Molecular dynamics (MD) simulations have emerged as a powerful tool to study the complex behavior of molecular systems. 
However, despite their utility, MD simulations face significant challenges in efficiently exploring the vast conformational space of macromolecules and capturing rare events that occur on time scales beyond the reach of conventional simulation methods.
Rare events, such as protein folding, ligand binding, chemical reactions, and large-scale conformational changes, play a pivotal role in numerous biological and chemical phenomena \cite{chipot2007free,peters2017reaction}. 
These events often involve transitions between multiple metastable states separated by high free energy barriers, such that the time scales associated with rare events can be orders of magnitude longer than those accessible to standard MD simulations, rendering direct observation impractical with reasonable computational resources.
Consequently, accurately characterizing the kinetics and thermodynamics of such rare events remains an important challenge.

To facilitate the efficient sampling of rare events, researchers have developed a diverse array of enhanced sampling techniques \cite{henin2022enhanced}, which bias the simulations toward regions of interest or lower energy barriers.
These techniques can roughly be divided into two classes: one that focuses on the sampling of transitions by manipulating the initial conditions\cite{chandler1978statistical,bolhuis2002transition,swenson2018openpathsampling} and the other that introduces a bias term to the potential energy surface (PES).
In this work, we concentrate on the latter class of algorithms, which encompasses a broad spectrum of methodologies, including umbrella sampling \cite{torrie1977nonphysical}, metadynamics \cite{laio2002escaping}, and adaptive biasing force \cite{darve2001calculating,comer2015adaptive}. 
All of them have in common that they require the selection of one or several collective variables (CVs), special degrees of freedom that describe the rare event of interest, such that biasing along these CVs will greatly accelerate the observation of the rare transitions.

For small molecular systems, it is often possible to choose a CV based on chemical intuition, e.g., the distance between two atoms whose bond is broken or formed during a chemical reaction.
In contrast, when dealing with complex high-dimensional systems, it is virtually impossible to choose a CV based on a physical basis.
As an example, the dissociation of table salt in water is insufficiently described by the sodium--chloride distance, as the surrounding water molecules play an important role \cite{geissler1999kinetic,wang2024local}.
In such cases, data-driven approaches can be and have been leveraged to identify a coordinate suitable for enhanced sampling \cite{mendels2018collective,sultan2018automated,wang2019past,bonati2020data,sun2022multitask,sipka2023constructing,bonati2023unified}.
Machine learning techniques have been employed to map high-dimensional and general descriptors, such as C$_\alpha$--C$_\alpha$ distances or contact maps, to lower-dimensional CVs utilizing available data, often obtained from metastable states through unbiased simulations. 
\citet{sultan2018automated} proposed using distances to the decision boundary, predicted from binary classifier models, as CVs describing transitions between two states. 
However, learning the decision boundary between states may not suffice for effective CVs in enhanced sampling, as thorough exploration within metastable states is also crucial. 
Deep Targeted Discriminant Analysis (Deep-TDA), as proposed by \citet{trizio2021enhanced}, addresses this challenge by fitting each known state into Gaussian distributions with predefined target mean and variance. 
Consequently, the model learns the meaningful variances within and between states.  
It has been effectively applied to chemically significant and intriguing processes such as enzymatic catalysis \cite{das2023and} and sulfur polymerization \cite{yang2024structure}.

\begin{figure*}[!ht]
\centering
    \includegraphics[width=0.75\textwidth]{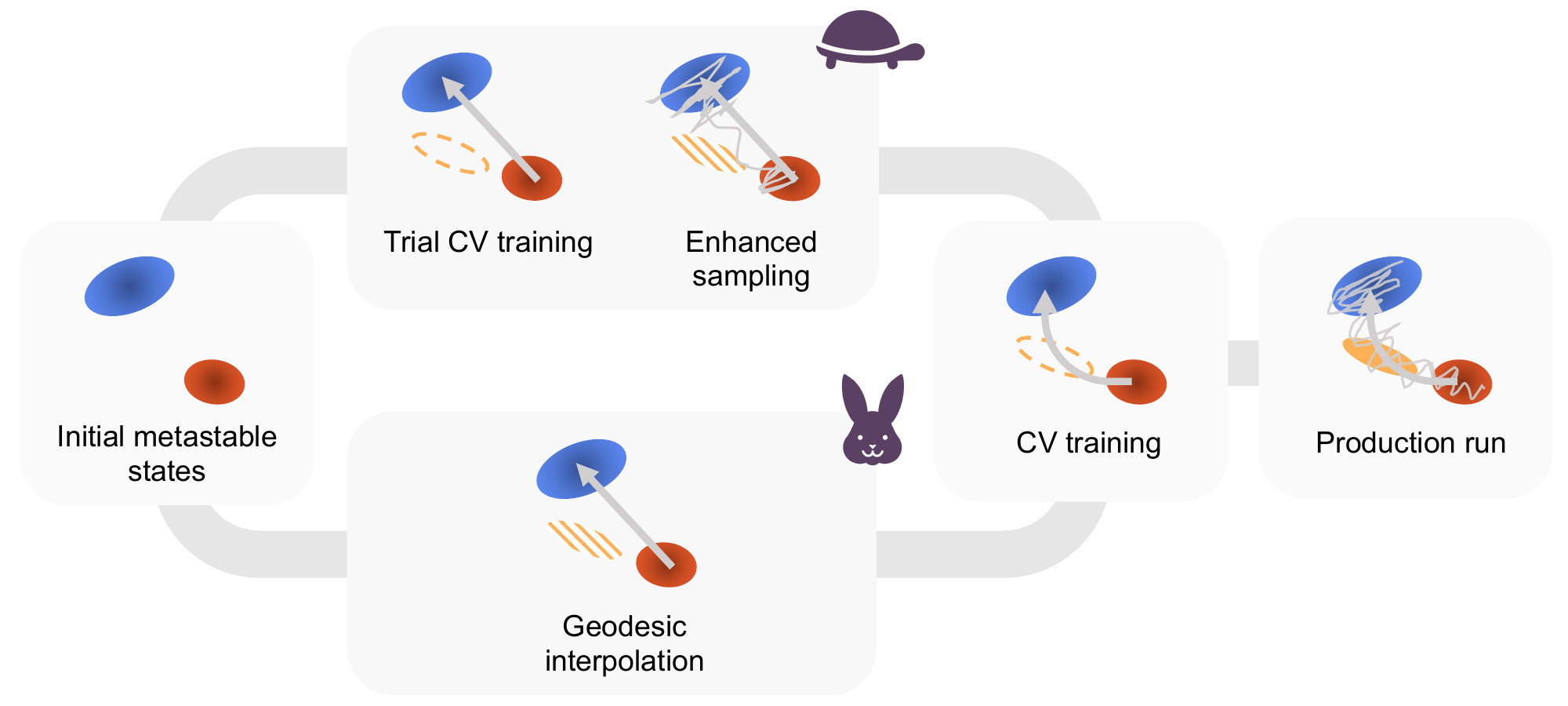}
    \caption{
    Depiction of a common pipeline for data-driven CVs.
    Initially, only data from metastable states are available (left).
    Long and costly production runs are meaningful only when performed with a reliable CV (right).
    Top (previous methods): A trial CV is trained and then iteratively improved through enhanced sampling simulations, which generate more data of the transition between the metastable states.
    Bottom (ours): Geodesic interpolations are used to create synthetic TSE data, from which the CV can be trained in one shot, obviating the need for an iterative procedure.
    }
    \label{fig:Fig1}
\end{figure*}

However, machine-learned models of CVs are data-hungry, and in the absence of information about the transition state ensemble (TSE), such data-driven CVs may not be optimal.
Hence, we are faced with a chicken-and-egg problem; we need enough samples of the rare event in order to obtain good CV models that are required to perform enhanced sampling of said event.
To circumvent this problem, recent methods, such as Deep Time-Lagged Independent Component Analysis (Deep-TICA) \cite{bonati2021deep} and Transition Path Informed Deep-TDA (TPI-Deep-TDA) \cite{ray2023deep}, adopt an iterative approach as illustrated in Figure~\ref{fig:Fig1}, where they first perform enhanced sampling with suboptimal CVs to collect initial trajectories, including a few transitions, and then use those samples to improve the CV.
While such methods have been proven to reduce the total simulation time required to obtain the target observable, the initial simulation for the data collection would have to be long enough to observe enough transitions. 
In complex systems, this step alone could be quite costly.

While classifier-based methods only utilize the state labels of each data point, in certain cases, we can obtain more informative labels that correlate with the reaction progress, enabling us to train a regressor model instead. 
Recently, \citet{france2024data} introduced a regression approach based on committor probabilities and \citet{lazzeri2023molecular} derived an algorithm for the unbiased Boltzmann weights from path sampling simulations. 
Additionally, \citet{kang2024computing} devised a way to learn the committor on the fly during iterations of enhanced sampling simulations.
However, a drawback of these approaches is their computational cost, as they require iterations of shooting simulations per data point or enhanced sampling simulations to compute the committor probability.
Even a recent CV-free approach, as introduced by \citet{sipka2023differentiable}, also requires resampling of the transitions between the metastable states to iteratively improve the biasing potential, thus it also has a high learning cost prior to production runs.

Recently, geodesic interpolation methods have been developed to interpolate approximately the transition path between two molecular conformations through purely geometric means.
\citet{zhu2019geodesic} proposed a Riemannian manifold based on a physics-inspired metric that effectively constructs a coarsened approximation of the PES.
The geodesic interpolations on this manifold closely resemble the minimum energy paths (MEPs) on the PES.
They demonstrated that these synthetic MEPs could serve as initial paths for subsequent nudged elastic band (NEB) optimization \cite{mills1995reversible}, ultimately bringing the initial MEP guess closer to the ground truth.
\citet{diepeveen2023riemannian} introduced a metric specifically designed to accommodate large-scale protein conformational changes and developed an algorithm that renders geodesic interpolation computationally feasible.
They additionally demonstrated that these interpolations closely resemble MD trajectories.
While initially proposed as a method for analysis and visualization of protein structural ensembles, we demonstrate its application to enhanced sampling.

To significantly reduce the computational cost of training effective CV models, we propose a strategy for simulation-free data augmentation.
Utilizing the Riemannian manifold for protein conformations proposed by \citet{diepeveen2023riemannian}, we generate geodesic interpolations, which resemble protein folding transitions remarkably well.
We show that augmenting training data with these interpolations improves the sampling even without the true transition state samples.
Another key advantage of these interpolations is that the interpolation parameter $t \in [0, 1]$ represents the progress of the transition.
This opens the door to supervised learning of protein folding in a regression setting, providing the model with richer information compared to discriminant analysis-based methods.
Our main contributions are:
\begin{itemize}
\item We propose an effective, simulation-free data augmentation strategy for CV learning in a protein folding context that significantly reduces the need for expensive simulations.    
\item We propose a novel regression-based learning scheme for CV models, where we exploit the progress parameter of geodesic interpolations as a supervisory signal. Our benchmark reveals that the regressor shows similar or better performance compared to the discriminant analysis method.  
\end{itemize}

\section{Methods}
\label{Methods}
\subsection{Geodesic interpolation of protein conformations}

Here, we leverage the computationally feasible technique for approximating geodesics on Riemannian manifolds that mimic protein energy landscapes introduced by \citet{diepeveen2023riemannian} to generate synthetic transition state data by performing geodesic interpolations between metastable protein states.
We summarize the geodesic interpolation approach below.
For rigorous notions of Riemannian geometry, we refer to the original work.\cite{diepeveen2023riemannian}

The heavy atom positions (all atoms but hydrogen) of a protein structure are embedded in a point cloud manifold $\mathcal{M} \subset \mathbb{R}^{n \times 3}/\mathrm{E}(3)$ where points do not overlap and are neither collinear nor coplanar.
We equip the manifold with a distance metric $w : \mathcal{M} \times \mathcal{M} \to [0, \infty)$ defined as
\begin{align}
    w(\mathbf{X}, \mathbf{Y}) = \sqrt{
        \sum_{i<j} \left( \log \frac{\left\Vert \mathbf{x}_i - \mathbf{x}_j \right\Vert}{\left\Vert \mathbf{y}_i - \mathbf{y}_j \right\Vert} \right)^2 + \left( \log \frac{\det S_{\mathbf{X}}}{\det S_{\mathbf{Y}}} \right)^2
    },
    \label{eq:metric}
\end{align}
where $\mathbf{X}, \mathbf{Y} \in \mathcal{M}$ are two different structures, $\mathbf{x}_i, \mathbf{y}_i \in \mathbb{R}^3$ are the positions of the $i$-th atoms in $\mathbf{X}$ and $\mathbf{Y}$, respectively, and $S_\mathbf{X} = \frac{1}{n} \sum_{i=1}^n (\mathbf{x}_i - \bar{\mathbf{x}}) (\mathbf{x}_i - \bar{\mathbf{x}})^\top \in \mathbb{R}^{3 \times 3}$ is the gyration tensor of $\mathbf{X}$.
The implication of the metric in eq~\eqref{eq:metric} is that a change in a small pairwise distance is more significant than a change of the same magnitude in a large pairwise distance. 
In other words, a perturbation of the Euclidean distance between two atoms affects the distance on the manifold more strongly when the atoms were originally close by. 
This aligns with the physical intuition that the important interactions between the particles in a system are mostly local, and the perturbations in the interactions correspond to changes in the energy of the system.

\begin{figure}[!ht]
    \includegraphics[width=0.8\textwidth]{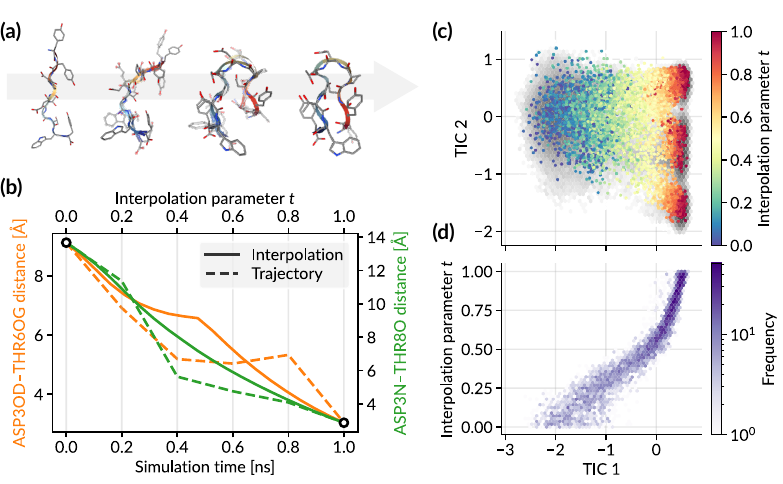}
    \caption{
    Comparison of the unfolded--folded transition observed in a long reference trajectory and the corresponding transition generated through geodesic interpolation of the end points.
    \textbf{(a)} Overlay of interpolated structures (opaque) and the reference structures (transparent). The initial and final structures are identical.
    \textbf{(b)} Evolution of the donor--acceptor distances for the two key hydrogen bonds observed in the folded state, in respective colors.
    The dashed lines are the reference transition as a function of time, and the solid lines are the interpolated conformations as a function of parameter $t$.
    \textbf{(c)} Overlay of 5,000 interpolated samples with $t$ uniformly sampled over the range of [0, 1], displayed on the first two slow modes from TICA. The projection of the reference unbiased trajectory is shown in gray in the background.
    Low TIC 1 values (left side) correspond to the unfolded basin, while high values (right side) correspond to the folded basin.
    \textbf{(d)} Interpolation parameter $t$ correlates with the progress along the slowest mode (TIC 1), which describes the transition.
    }
    \label{fig:Fig2}
\end{figure}

Now, the interpolation $\mathbf{Z}$ between two structures $\mathbf{X}, \mathbf{Y} \in \mathcal{M}$ with weights $(t, 1-t)$ is given by the $w$-geodesic $\gamma_{\mathbf{X},\mathbf{Y}}^w:[0, 1] \to \mathcal{M}$ defined as
\begin{align}
    \gamma_{\mathbf{X}, \mathbf{Y}}^w(t) = \operatornamewithlimits{argmin}_{\mathbf{Z} \in \mathcal{M}} \left( \frac{1-t}{2} w(\mathbf{X}, \mathbf{Z})^2 + \frac{t}{2} w(\mathbf{Z}, \mathbf{Y})^2 \right)\ ,
    \label{eq:geodesic}
\end{align}
so that the interpolated structure $\gamma_{\mathbf{X}, \mathbf{Y}}^w(t)$ is a minimizer of the inversely weighted sum of squared distances to the two structures.
See Figure~\ref{fig:Fig2}(a) for an illustrative example of interpolated structures.
 
\subsection{Leveraging the interpolation parameter as an indicator of reaction progress}

Methods based on discriminant analysis heavily rely on accurate state labels, as the loss functions are tailored to maximize the differentiation between these states. 
As reported in \citet{ray2023deep}, training the CV model with reduced overlap between states can significantly enhance its performance.  
However, distinguishing metastable state samples from TSE samples becomes challenging in situations where our understanding of the system is limited.
For TPI-Deep-TDA, the remedy for this problem is filtering out metastable configurations from the TSE data with the initial TDA model.\cite{ray2023deep}
The interpolation progress parameter $t$ can be considered as a proxy for the reaction progress, as illustrated in Figure~\ref{fig:Fig2}. 
In this work, we propose training a regressor CV leveraging this additional source of information as an option to go beyond discrete state labels. 

\begin{figure}[!ht]
\centering
    \includegraphics[width=0.6\textwidth]{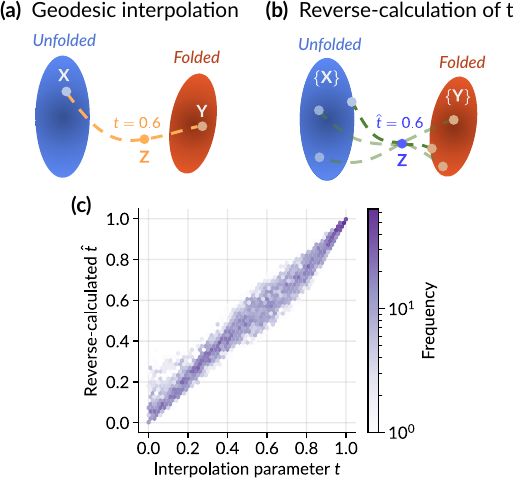}
    \caption{\textbf{(a)} Scheme for geodesic interpolation between unfolded and folded conformations.
    The interpolated structure $\mathbf{Z}$ corresponding to interpolation parameter $t$ is obtained using the geodesic in eq~\eqref{eq:geodesic}.
    \textbf{(b)} Procedure to estimate interpolation parameter $\hat{t}$ of a given intermediate conformation and sets of unfolded and folded conformations, which is given by a ratio of the minimum distances to unfolded and folded conformations (eq~\eqref{eq:reverse_cal_t}).
    \textbf{(c)} Parity plot comparing the true interpolation parameter $t$ with the reverse-calculated $\hat{t}$ for 5,000 interpolations, with $t$ uniformly sampled over the range [0, 1].
    }
    \label{fig:Fig3}
\end{figure}

As shown in eq~\eqref{eq:geodesic} it is always possible to find a structure $\gamma_{\mathbf{X}, \mathbf{Y}}^w(t)$ that interpolates between two given samples from the metastable basins.
The interpolated structures are generated from the ensemble of metastable state structures.
Since the formulation in eq~\eqref{eq:geodesic} involves a single pair of structures, the parameter $t$ implicitly depends on the sampled end point structures.
However, we can ``inversely'' estimate the parameter $\hat{t}$ given the sets of metastable state structures as a ratio of the minimum geodesic distance to each metastable state as
\begin{align}
    \hat{t}(\mathbf{X}\,\vert\,\mathcal{U},\mathcal{F}) = \frac{\min_{\mathbf{U} \in \mathcal{U}}(w(\mathbf{X},\mathbf{U}) 
    )
    }{\min_{\mathbf{U} \in \mathcal{U}}(w(\mathbf{X},\mathbf{U}) 
    )
    +\min_{\mathbf{F} \in \mathcal{F}}(w(\mathbf{X},\mathbf{F}) 
    )
    } \ ,
    \label{eq:reverse_cal_t}
\end{align}
where $\mathbf{F}$ and $\mathbf{U}$ are samples from the set of folded and unfolded conformations, $\mathcal{F}$ and $\mathcal{U}$, respectively.
See Figure~\ref{fig:Fig3} for a depiction of both eqs.~\eqref{eq:geodesic} and~\eqref{eq:reverse_cal_t}.
In our experiments, we compare the $t$ and $\hat{t}$ to demonstrate that the end point dependence of the parameter $t$ is not strong and could be regarded as an approximate function of metastable state ensembles.
Furthermore, given an intermediate structure without labels, our reverse-calculation scheme enables ``labeling'' of the structures according to the approximate reaction progress without the knowledge of the actual transition path.

\subsection{Chignolin as a model system}

To demonstrate the capabilities of our enhanced sampling method for protein simulations, we utilize the chignolin variant CLN025 as a benchmark system. 
Despite its small size of 10 residues and 166 atoms, it folds into a stable $\beta$-hairpin structure.
Simulating this folding process poses a nontrivial sampling challenge that necessitates very long simulation times in unbiased simulations. 
Hence, chignolin is recognized as a popular benchmark system for the evaluation of enhanced sampling methods. 
In this work, the 106-$\mu$s long unbiased simulation trajectory of chignolin \cite{lindorff2011fast} serves as reference data for assessing the accuracy and efficiency of our CVs.
An average folding time of 0.6~$\mu$s was reported in the reference simulation for chignolin \cite{lindorff2011fast}.

\subsection{Transition-focused data augmentation}

While simulations often generate abundant data within metastable states, capturing the rare transition state is challenging \cite{kang2024computing}. 
Incorporating this scenario, we aim to train a robust CV model by utilizing the abundant metastable state data and synthetic transition state data.  

We begin by performing short (50~ns) MD simulations, one for each metastable state (folded and unfolded), generating 5,000 data points per state.
We imposed restraints based on the C$_\alpha$ root-mean-squared deviation (RMSD) from the reference folded conformation, such that no transitions (folding or unfolding) can occur.
Note that, while the TPI-Deep-TDA study \cite{ray2023deep} sampled configurations every 1~ps, we sampled every 10~ps, corresponding to a 10-fold reduction in the number of metastable state data points compared to their study.
In our opinion, closer temporal spacing between the sampled geometries would not improve the performance of machine learning collective variables (ML CVs) due to the similarity of the geometries.
Then, to mimic transitions, we perform 5,000 geodesic interpolations between the metastable state geometries using heavy atom coordinates.
We control the distribution of the synthetic TSE by sampling $t$ from a target distribution.
Considering the abundant data from the metastable states, we sample $t$ from a Gaussian distribution centered at 0.5 to produce more conformations closer to the transition state, as values of $t$ near 0 or 1 correspond to the samples belonging to the metastable states.
As depicted in Figure~\ref{fig:Fig2}(a), geodesic interpolation from an unfolded (left) to a folded (right) state yields structures (middle, opaque) resembling those from MD simulations (middle, transparent). 
Detailed information regarding the geodesic interpolations is provided in Section~S1.2 of the Supporting Information (SI).

\subsection{Machine learning collective variable models}

ML CVs are designed to map high-dimensional structural descriptors onto lower-dimensional CVs.
For all models, we featurize the inputs with all 45 pairwise contacts between ten C$_\alpha$ atoms.

Given class labels for metastable state samples, CV learning with targeted discriminant analysis (TDA) loss, introduced by \citet{trizio2021enhanced}, would be a natural starting point.
In the case of one-dimensional CVs, TDA assumes that there exist a nonlinear transformation such that the marginal Boltzmann distribution along the CV forms a mixture of $N_s$ Gaussians with user defined means and variances.
The TDA loss is defined as shown in eq~\eqref{eq:TDA}, where $N_s$ represents the total number of states, ${\mu}_{k}$ and ${\sigma}_{k}$ are mean and standard deviation of CV model predictions for the training data with ground truth state label $k$, $\bar{\mu}_{k}$ and $\bar{\sigma}_{k}$ denote the target Gaussian mean and standard deviation corresponding to state $k$, and $\alpha$, $\beta$ are hyperparameters.
\begin{equation}
    \mathcal{L}_\text{TDA}=\alpha \sum_{k=1}^{N_s}\left(\mu_{k}-\bar{\mu}_{k}\right)^2+\beta \sum_{k=1}^{N_s}\left(\sigma_{k}-\bar{\sigma}_{k}\right)^2
    \label{eq:TDA}
\end{equation}
This loss function enforces the distribution of model predictions for the training data from each metastable state closely matches the corresponding target Gaussian distribution.

We train multilayer perceptron (MLP) models with the TDA loss, both with and without interpolated samples. 
The model trained solely on the two metastable states ($N_s=2$) is denoted as \textbf{TDA}, while the one trained on the two metastable states along with the synthetic transition state generated from geodesic interpolation is labeled as \textbf{TDA\textsubscript{geo}} ($N_s=3$). 

In comparison to TDA models trained on class labels, we also train MLP regression models using the interpolation parameter $t$ as labels, optimized with mean squared error (MSE) loss. 
The unfolded and folded configurations are assigned target labels of 0 and 1, respectively.
The regression model trained on both metastable states and the synthetic TSE data is referred to as \textbf{Reg\textsubscript{geo}}.
For both architectures, we compare augmenting them with the full set of interpolated structures and the subset of those structures that, according to the two-state TDA, do not belong to either metastable state (denoted as unfiltered and filtered, respectively). 

More details on the model training are described in SI, Section~S1.1.
 
\subsection{Enhanced sampling and result analysis}

For each ML CV, we perform five 1-$\mu$s well-tempered metadynamics extended-system adaptive biasing force (WTM-eABF) \cite{fu2019taming} simulations initiated from distinct unfolded configurations.
WTM-eABF enhances the sampling along the chosen CV by effectively flattening the energy landscape with the combination of the extended-system adaptive biasing force (eABF, \citet{fu2016extended}) and well-tempered metadynamics (WTM, \citet{barducci2008well}).
In ABF, the negative running estimate of average force along the CV is applied to adaptively shear off the barrier, while in WTM, the adaptive potential is stacked along the CV to fill up the energy landscape to the barrier level \cite{chipot2023free}.
We evaluate the performance of our ML CVs by analyzing the convergence of the estimated reaction free energy of folding and the resulting potential of mean force (PMF).

The PMF $A(s)$ is obtained as a function of CV value $s$ from the biased simulations via multistate Bennett acceptance ratio (MBAR) analysis \cite{shirts2008statistically,hulm2022statistically}.
Integrating the PMF over each metastable basin allows for the computation of the free energy difference between the folded and unfolded states \cite{dietschreit2022obtain}:
\begin{equation}
\Delta F = -k_\textrm{B}T \log \left( \frac{\int_\text{folded} \exp (-A(s) / k_\textrm{B}T) \, \mathrm{d}s}{\int_\text{unfolded} \exp (-A(s) / k_\textrm{B}T) \, \mathrm{d}s} \right).
\label{eq:deltaF}
\end{equation}
We report the mean and standard deviation of all five runs except a single run from \textbf{TDA} simulations which showed abnormal detachment of the CV and the fictitious particle position $\lambda$, as shown in the time evolution of CV and $\lambda$ in SI Figure~S2.

In addition to $\Delta F$, we introduce the mean absolute error (MAE) of the PMF, defined as follows:
\begin{align}
    \text{MAE}(A, A_\text{ref}) = \frac{\int \left\vert A(s) - A_\text{ref}(s) \right\vert \mathbb{I}\left[A_\text{ref}(s) < A_\text{thres}\right] \, \mathrm{d}s}{\int \mathbb{I}\left[A_\text{ref}(s) < A_\text{thres}\right] \, \mathrm{d}s},
    \label{eq:pmf_mae}
\end{align}
where $\mathbb{I}[\cdot]$ is an indicator function and $A(s)$ and $A_\text{ref}(s)$ are PMFs for the same CV obtained from the five (four for \textbf{TDA}) enhanced sampling runs and the long reference trajectory, respectively. 
The PMFs are aligned by setting minimum values to zero.
This metric quantifies the average difference between the two PMFs within relevant CV regions characterized by reference PMF values lower than the threshold $A_\text{thres} = 25$~kJ/mol.
Alongside capturing the PMF difference within metastable basins, as indicated by $\Delta F$, the PMF MAE also assesses deviations from the reference PMF in transition regions.

\section{Results and Discussion}
\label{Experiments}

\begin{figure*}[!ht]
    \includegraphics[width=\textwidth]{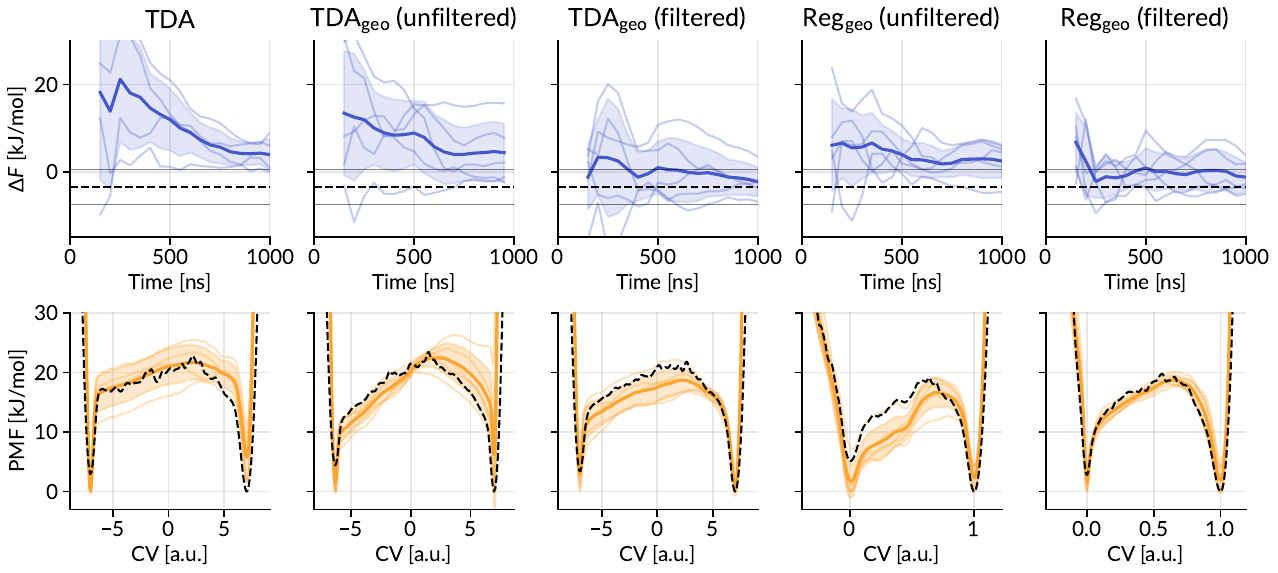}
    \caption{Convergence of the free energy difference ($\Delta F$) between folded and unfolded state and the PMF at the end of the simulation as sampled with WTM-eABF using each CV. 
    Top panel: Evolution of the $\Delta F$ estimate over the course of the trajectory.
    Horizontal solid lines indicate a $\pm$ 4~kJ/mol range (chemical accuracy) around the $\Delta F$ value obtained from the long unbiased reference runs (dashed line).
    Bottom panel: Comparison of reference PMF obtained by projection of reference data (dotted line) and those obtained from 1~$\mu$s WTM-eABF simulations with the given CV.
    The shaded areas represent the standard deviation of the independent simulations, and the solid line is the mean.}
    \label{fig:DeltaF_PMF}
\end{figure*}

\subsection{Geodesic interpolation resembles state-state transitions}

In this section, we investigate the properties of the interpolated data and the parameter $t$.
The intermediate structures generated by using geodesic interpolation smoothly interpolate between the unfolded and folded structures, with a close resemblance to the unbiased reference trajectory over simulation time, as visualized in Figure~\ref{fig:Fig2}(a).
The donor--acceptor distances of the two key hydrogen bonds between Asp3--Thr6 and Asp3--Thr8 are also smoothly interpolated, following the trend of those from the unbiased trajectory (Figure~\ref{fig:Fig2}(b)).

Time-lagged independent component analysis (TICA) is a dimensionality reduction technique that effectively diagonalizes a time-lagged correlation matrix \cite{molgedey1994separation,perez2013identification}.
Where principal component analysis (PCA) returns eigenvectors that are linear combinations of the input features that maximize the corresponding variance, TICA returns the combinations with the slowest motions.
The first time-lagged independent component (TIC) represents the slowest collective motions of the chignolin system, capturing the unfolded--folded transition, and the second component resolves the three clusters within the folded state. 
Figure~\ref{fig:Fig2}(c) shows that the interpolated structures and the reference transition paths have very similar projections on the first two TICs.
Furthermore, as shown in Figure~\ref{fig:Fig2}(d), the interpolation parameter $t$ has a strong correlation with the slowest mode of the system (TIC 1), indicating that $t$ is in fact a useful descriptor of the progress of chignolin folding.
Finally, the comparison of $t$ and $\hat{t}$ in Figure~\ref{fig:Fig3}(c) demonstrates that the two values are well aligned with each other.
This indicates that the end point dependence of the parameter $t$ is weak, allowing it to be interpreted as a measure of the transition progress between two metastable state ensembles rather than being linked to specific samples of them.

For chignolin, interpolating 5,000 frames took approximately 40 min, which is about half the time of each 50 ns unbiased metastable state simulation needed to extract the same number of frames.
Further details regarding computation time and scaling results are provided in the SI, Section~S3.

Compared to the physical TSE, such as those derived from the on-the-fly probability enhanced sampling (OPES) flooding approach \cite{ray2022rare,ray2023deep}, the pseudo-TSE generated from the geodesic interpolation is advantageous in being simulation parameter-free, faster, and allowing for controllable generation (by means of $t$) and facilitating the generation of decorrelated samples.
However, this pseudo-TSE is purely geometric in nature and is not guaranteed to represent the physical one.
On the other hand, utilizing enhanced sampling simulations to generate samples produces physically realistic structures, but is slower and requires the system to escape the metastable basin.
It also results in correlated samples, i.e., one run only produces a single path exploring the TSE, which then quickly goes over into the other metastable basin.
Additionally, it necessitates some simulation parameter settings, such as the expected barrier height.
Despite these differences, both methods share the benefit of being parallelizable, enabling efficient computation across multiple processors.
We expect that the choice of method may depend on the user's familiarity with the enhanced sampling protocol and the available computational budget for simulations.

\begin{table}[!ht]
\caption{Mean free energy of folding ($\Delta F$) obtained from 1-$\mu$s WTM-eABF simulations (Figure~\ref{fig:DeltaF_PMF}) with the corresponding reference value from the long unbiased simulation, and the mean absolute error of the potential of mean force (PMF MAE, eq~\eqref{eq:pmf_mae}) for each CV.
Values in parentheses are the standard deviation from four (TDA) or five (others) independent simulations.
All values are reported in units of kJ/mol.
}
\label{tab:results}
\begin{center}
\begin{tabular}{lrrr}
\toprule
\multicolumn{1}{c}{CV model} & \multicolumn{1}{c}{$\Delta F$} & \multicolumn{1}{c}{$\Delta F_\text{ref}$} & \multicolumn{1}{c}{PMF MAE} \\
\midrule
\textbf{TDA} & $3.92$ ($3.73$) & $-3.59$ & $4.44$ ($1.09$) \\
\textbf{TDA\textsubscript{geo}} (unfiltered) & $4.37$ ($6.67$) & $-3.56$ & $6.29$ ($1.57$) \\
\textbf{TDA\textsubscript{geo}} (filtered) & $-2.31$ ($3.32$) & $-3.58$ & $3.53$ ($0.98$) \\
\midrule
\textbf{Reg\textsubscript{geo}} (unfiltered) & $2.45$ ($3.88$) & $-3.64$ & $3.77$ ($1.36$) \\
\textbf{Reg\textsubscript{geo}} (filtered) & $-1.24$ ($3.41$) & $-3.59$ & $1.93$ ($0.51$) \\
\bottomrule
\end{tabular}
\end{center}
\end{table}

\subsection{Data augmentation improves ML CV}

\begin{figure*}[!ht]
    \includegraphics[width=\textwidth]{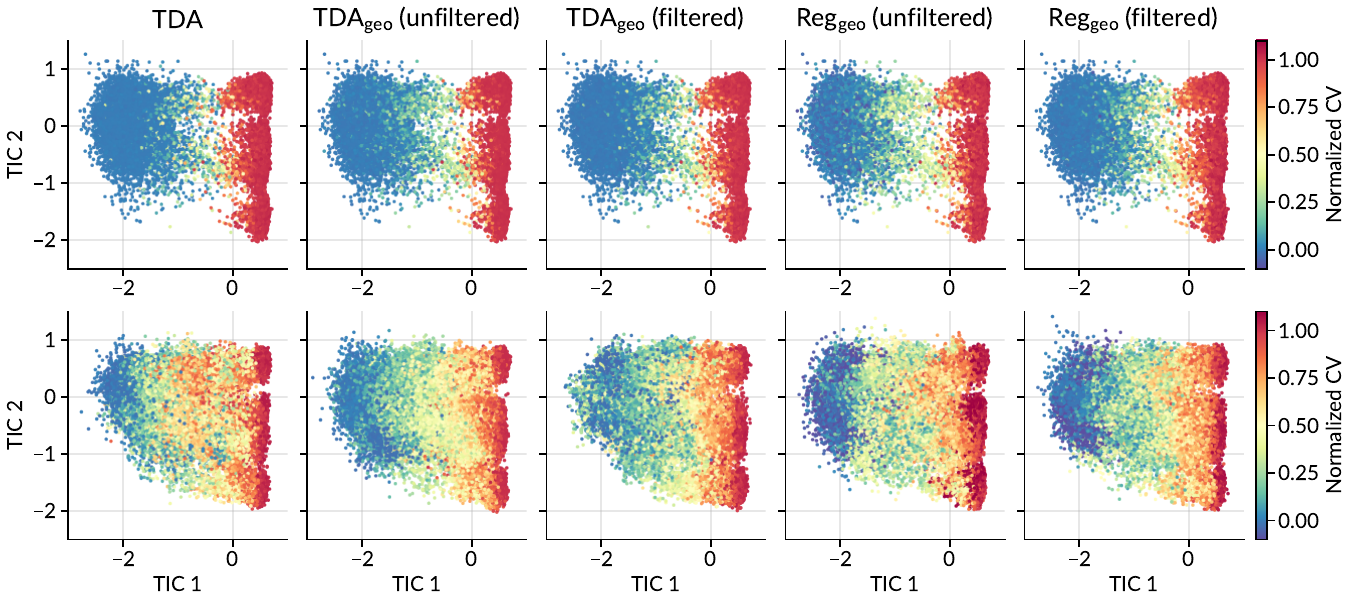}
    \caption{Projections of the conformations from the unbiased reference trajectory (upper rows) and WTM-eABF enhanced sampling trajectory (lower rows) using different CV models onto the first two time-lagged independent components, colored based on the normalized CV value. The CV values were scaled such that the PMF minima of the unfolded and folded basins correspond to 0 and 1, respectively.}
    \label{fig:tica_cv}
\end{figure*}

For each of the statistically independent 1-$\mu$s WTM-eABF enhanced sampling simulations, with the first 100~ns discarded as equilibration time, we obtained the PMF and calculated the free energy of folding ($\Delta F$, eq~\eqref{eq:deltaF}) as a function of simulation time.
The convergence of $\Delta F$ and the final PMFs obtained from the enhanced sampling runs are reported in Figure~\ref{fig:DeltaF_PMF}.
We further compare the sampled PMFs with those obtained from the reference trajectory by investigating the difference between $\Delta F$ and the reference value $\Delta F_\text{ref}$ and evaluating the PMF MAE (eq~\eqref{eq:pmf_mae}), as collected in Table~\ref{tab:results}.

First, the $\Delta F_\text{ref}$ values obtained from the projection of the reference trajectory onto the CV lie within a very narrow range of $-3.64$ to $-3.56$ kJ/mol.
This consistency indicates that all of our trained CVs can correctly distinguish between the unfolded and folded equilibrium conformations in the reference trajectory.

Figure~\ref{fig:DeltaF_PMF} and Table~\ref{tab:results} show that models trained with data from the metastable states augmented with filtered pseudo-TSE data perform the best (\textbf{TDA\textsubscript{geo}} (filtered) and \textbf{Reg\textsubscript{geo}} (filtered)).
They are the only models that, on average, successfully approach the reference value of $\Delta F$ within the chemical accuracy threshold of 4~kJ/mol. 
Surprisingly, all other models predict the unfolded state to be more stable. 
Further, using all 5,000 pseudo-TSE configurations to train a three-state TDA model, \textbf{TDA\textsubscript{geo}} (unfiltered), shows no improvement compared to the two-state \textbf{TDA} model. 
The reason for this likely being the significant overlap of the distribution of structures generated by the geodesic interpolation and those from the metastable state simulations. 
Both \textbf{TDA\textsubscript{geo}} and \textbf{Reg\textsubscript{geo}} show faster convergence in terms of the average $\Delta F$ value, even if it is to the wrong value, than \textbf{TDA} without data augmentation.
Notably, the CVs trained on the filtered data quickly converge to the reference $\Delta F$ in less than 500~ns and demonstrate a very good reproduction of the reference PMF.

The enhancement in performance from the data augmentation can be attributed to the CV model's ability to generalize its classification ability to off-equilibrium conformations observed in the enhanced sampling runs and accurately identify the transition progress.

\subsection{Analysis of CV model predictions for sampled configurations}

\begin{figure*}[!ht]
    \includegraphics[width=\textwidth]{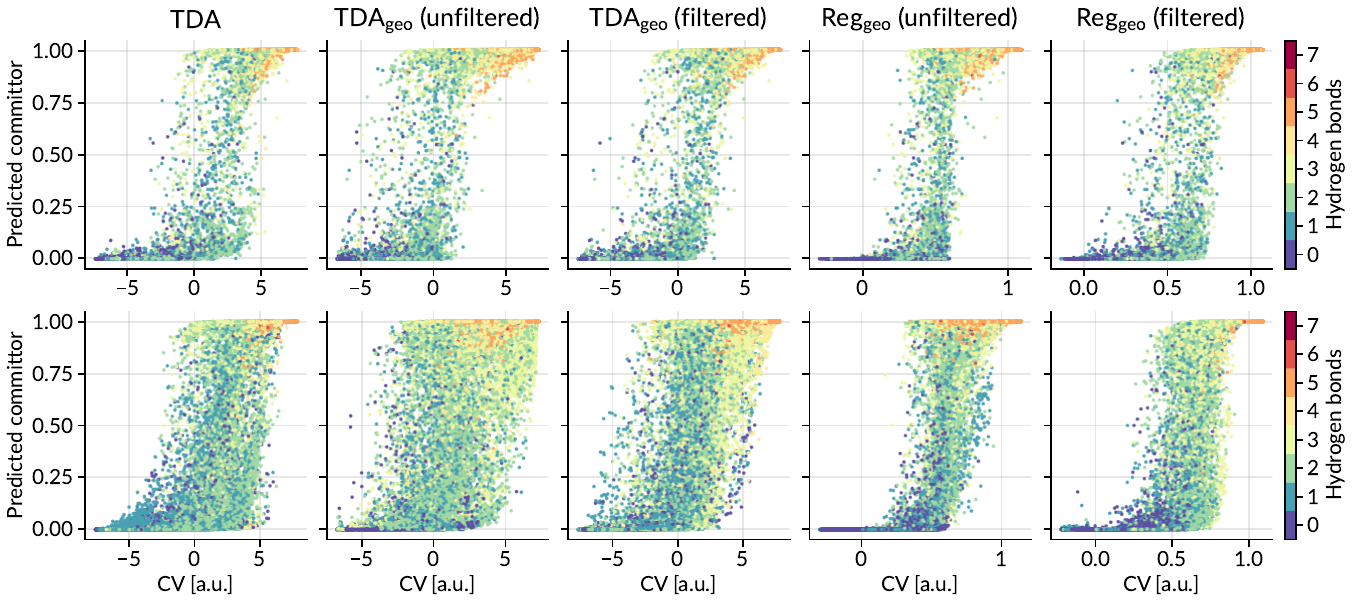}
    \caption{
    CV and predicted committor values (from \citet{kang2024computing}, subtracted from one to match the transition direction) for the conformations from the unbiased reference trajectory (upper rows) and WTM-eABF enhanced sampling trajectory (lower rows).
    Each point is colored according to the number of native hydrogen bonds formed in the corresponding conformation, as defined in Figure~S1.}
    \label{fig:cv_committor}
\end{figure*}

To better understand the behavior of the different ML CVs, we show two low-dimensional projections of the data: TIC~1--TIC~2 (Figure~\ref{fig:tica_cv}) and CV value--committor prediction (Figure~\ref{fig:cv_committor}).
First, in Figure~\ref{fig:tica_cv}, we have projected all conformations of the reference trajectory, as well as the trajectories generated by the different enhanced sampling runs, onto the space spanned by the first two TICs (TIC 1 and TIC 2).
The TICA analysis was only performed on the long, unbiased reference simulation, i.e., the projection operation is the same for all subplots, and only the data varies.
The points were colored by scaled CV values (such that each CV is zero for the unfolded and one for the folded minimum on the PMF). 
Second, in Figure~\ref{fig:cv_committor}, we plotted the CV value of each configuration against its committor estimate, where the committor function was obtained in a data-driven manner in the recent study by \citet{kang2024computing}.
The points are colored by the number of hydrogen bonds formed in the native folded conformation, a physical observable that correlates well with the folding process of chignolin \cite{mckiernan2017modeling,sobieraj2022granger}.
A detailed illustration of the hydrogen bond formation in the chignolin folding process can be found in SI, Section~S1.3.

The first row of Figure~\ref{fig:tica_cv} shows almost identical plots for different CV models.
All ML CVs assign normalized CV values near one to the folded configurations (TIC 1 above 0) and much smaller values to the unfolded ones. 
From the TIC projections of the reference trajectory, chignolin has three connected folded state basins and the unfolded/folded transition occurs via two major channels that are connected to the two upper folded basins.
From the projections of all enhanced sampling simulations, the three folded state basins are clearly detected, but the exchange with the unfolded state occurs much more broadly.
Observing the coloring of the projections clarifies why \textbf{TDA} did not perform well compared to other CV models. 
Many configurations to which the model assigns a high CV value lie in the transition region or in the unfolded basin according to the first two TICs. 
A similar observation can be made for the regressor model trained on the unfiltered data, although not as strongly. 
In general, for none of the ML CVs we can see such a clear gradient in color from left to right as in the upper row.

The first row of Figure~\ref{fig:cv_committor} shows the TICA projection of the unbiased reference data, while the bottom row shows the data obtained from the respective enhanced sampling runs, similar to Figure~\ref{fig:tica_cv}.
Figure~\ref{fig:cv_committor} shows that the number of formed hydrogen bonds (as indicated by the color of the dots) correlates well with the committor value; however, for all models and all runs, there are configurations with no formed contacts but committor values indicating an almost folded state.
Note that we have used one minus the committor value from \citet{kang2024computing} such that it is zero for unfolded and one for folded conformations.
The upper row shows the expected sigmoidal curves for committor vs. CV, as the ML CVs change more gradually. 
For the unbiased geometries, all CV models show a clear separation of the extreme committor values (near zero and one), which is in agreement with the first row of Figure~\ref{fig:tica_cv} and explains the general agreement for the reference $\Delta F$ among all ML CV models.
However, one can observe a distinct broadening, especially for all TDA-type models in the bottom row of Figure~\ref{fig:cv_committor}. 
In order to observe a good estimation of the reaction free energy, a CV needs to clearly distinguish between the different metastable state, i.e., it needs to separate committor values of 0 and 1. 
Especially for the unfiltered \textbf{TDA\textsubscript{geo}} and \textbf{Reg\textsubscript{geo}} this is not the case. 
For \textbf{Reg\textsubscript{geo}} (unfiltered), one can see the overlap of extreme committor values for predicted CV values of around 0.5, which explains the noticeable lowering of the PMF in the transition region for this model in Figure~\ref{fig:DeltaF_PMF}.
The bottom row of Figure~\ref{fig:cv_committor} clearly explains why the two models trained on the TDA-filtered data perform better than those trained on unfiltered data.

\subsection{Current limitations of ML CVs}

All CVs presented in this work are machine learning models trained exclusively on data stemming from unbiased simulations and a pseudo-TSE, which in turn is also based on unbiased simulations.
The committor function is itself also an ML model, but since it was trained iteratively through enhanced sampling, it is assumed to be robust. 

While our best models were successful in finding correct $\Delta F$ and PMF, there still exists a question: How can we test whether simulations with ML CVs are trustworthy? All models studied here converged while training with low validation loss.
Further, using these CVs and the unbiased reference trajectory to compute $\Delta F$ consistently led to accurate estimations, let aside the fact that such reference data is not typically available in real-world scenarios. 
Yet, neither of these metrics could predict the actual performance of these models when used for enhanced sampling. 
This is a general problem of (black box) ML CVs. 
As accurate as they might be on easily classifiable data, we have little control over how they behave during enhanced sampling simulations where they will encounter many high-energy conformations.
Depending on the CV landscape and the direction of the CV gradient in relation to the minimum energy path, ML CV-based enhanced sampling simulations can potentially diverge significantly from the expected ensemble.  
In other words, we do not know beforehand how these models behave on out-of-distribution samples, a problem that usually does not exist for hand-picked physical degrees of freedom as poor as those might be.

As we currently do not have a satisfying answer to this problem, future work will have to find a way to increase the robustness of ML-based CVs in order to ensure the trustworthiness of the results obtained with them.
A possible avenue might be training multitask CVs as suggested by \citet{bonati2023unified}, e.g., one where an autoencoder framework is trained with a reconstruction loss combined with a TDA-type loss on the latent space. 
The reconstruction loss might introduce necessary robustness and is perfectly usable with synthetic, unlabeled data (lacking clear state labels) as provided by our geodesic interpolation method.
Another approach might involve the inclusion of true transition data, as done in TPI-Deep-TDA \cite{ray2023deep}, to guide the model in the correct direction.
However, this would not align with the fast, one-shot approach of this work.

\section{Conclusion}
\label{Conclusion}
We developed a scheme to train collective variables (CVs) from metastable state configurations using geodesic interpolation to generate a synthetic transition state ensemble (TSE).
We found that the generated TSE closely aligns with the reference transition configurations and provides additional information on the transition progress via the interpolation parameter $t$.
By sampling the parameter from a predefined distribution, we could control the distribution of the generated TSE over the transition progress.
Furthermore, we devised a method to reverse calculate the interpolation parameter, potentially enabling the incorporation of transition data obtained, e.g., by flooding simulations.
By leveraging the interpolation parameter label as a regression target for CVs, we demonstrated notable improvements in the enhanced sampling simulation results when compared to the CVs trained solely on metastable state conformations.
These results highlight the importance of incorporating transition state information into CV model training.
This not only enhances the ability to accurately distinguish between the metastable states, but also enables us to effectively capture the underlying dynamics of the system, as evidenced by the change in a number of hydrogen bonds along the CV axis.
We note that the behavior of ML CVs for out-of-distribution samples in enhanced sampling simulations could not be perfectly controlled with the use of pseudo-TSE data.

We emphasize that geodesic interpolation is not exclusively limited to protein conformations but can be generalized to a broader spectrum of rare events.
We anticipate that further development of this method would be applicable in various chemical events, ranging from study of folding pathways of larger proteins to modeling complex chemical reactions.

\section*{Code availability}
The code to reproduce this work is available on GitHub: \url{https://github.com/learningmatter-mit/geodesic-interpolation-cv}.
The PLUMED \cite{tribello2014plumed} input files are also available on PLUMED-NEST (\url{www.plumed-nest.org}), the public repository of the PLUMED consortium \cite{plumed2019promoting}, as plumID:24.014.

\begin{acknowledgement}
The authors acknowledge the MIT SuperCloud and Lincoln Laboratory Supercomputing Center for providing HPC resources that have contributed to the research results reported in this article.
The authors thank D. E. Shaw Research for sharing the reference trajectory data of chignolin (CLN025) in \citet{lindorff2011fast}.
We also want to thank Martin \v{S}\'{i}pka, Luigi Bonati, and Gianmarco Lazzeri for helpful discussions.
We acknowledge the anonymous reviewers for their valuable insights, including their suggestions to refs \cite{kang2024computing,bonati2023unified,lazzeri2023molecular,das2023and,yang2024structure}.
S.Y. is supported by Takeda Fellowship and Ilju Overseas Ph.D. Fellowship.
J.N. is supported by the Ronald A. Kurtz Fellowship.
J.C.B.D. is thankful for the support of the Leopoldina Fellowship Program, German National Academy of Sciences Leopoldina, grant number LPDS 2021-08. 
\end{acknowledgement}

\begin{suppinfo}
Details of the molecular dynamics and enhanced sampling simulations,
specifics of the data augmentation using geodesic interpolation and CV training,
structure-based analysis using hydrogen bonds present in the folded state,
additional simulation analysis, including the time evolution of CV and C$_\alpha$ TICA projections,
and analysis of computational costs
\end{suppinfo}

\bibliography{main}

\end{document}


\newpage
\tableofcontents
\newpage

\section{Computational Details}

\subsection{Simulation details}
\label{sec:simulation_details}

\paragraph{Molecular dynamics.}
Molecular dynamics simulations of chignolin (CLN025 variant) were performed with GROMACS 2023 \cite{abraham2015gromacs} patched with PLUMED 2.9.0 \cite{tribello2014plumed}.
The system was solvated in a cubic water box of size 39.61 $\times$ 39.61 $\times$ 39.61 \AA$^3$, and neutralized with two Na$^+$ ions.
The protein molecule was parametrized with the CHARMM22* force field \cite{piana2011robust} and the water molecules with the CHARMM-modified TIP3P model \cite{mackerell1998all} analogously to the reference simulation \cite{lindorff2011fast}.
Simulations were performed in the NVT ensemble.
The temperature was controlled with the velocity rescaling thermostat (\texttt{v-rescale}) \cite{bussi2007canonical} at 340~K. 
The equations of motion were integrated with the leap-frog algorithm with a time step of 2~fs.
The LINCS algorithm \cite{hess1997lincs} was used to constrain all bonds involving hydrogen atoms.

\paragraph{Unbiased simulations.}
Starting from the unfolded and folded conformations of chignolin, we prepared the simulation box as detailed in the previous paragraph, minimized the system energy, and performed equilibration for 10 ps.
After equilibration, we conducted a 50 ns MD simulation for each metastable state.
For the unfolded state MD, we applied a harmonic wall restraint with a spring constant of 500.0 kJ/mol/nm$^2$ to the C$_\alpha$ RMSD from the folded structure (PDB 5AWL) at 0.3 nm to prevent the system from folding.
The same harmonic restraint was applied in the opposite direction during the folded state MD to prevent unfolding.
We sampled five configurations from the unfolded state simulation to prepare five statistically independent initial configurations for the WTM-eABF enhanced sampling simulations.

\paragraph{Enhanced sampling parameters.}
We used the combination of \texttt{DRR} and \texttt{METAD} bias modules in PLUMED to perform WTM-eABF simulations and defined machine-learned collective variables using the \texttt{PYTORCH\_MODEL} module.
We used a \texttt{FRICTION} of 1.0 ps$^{-1}$, \texttt{TAU} of 10 ps for the DRR bias and \texttt{PACE} of 500 steps (1 ps), \texttt{HEIGHT} of 0.965 kJ/mol ($\approx$ 10 meV), \texttt{BIASFACTOR} of 12.765 for the well-tempered metadynamics bias.
We applied harmonic wall restraints to the fictitious particle position at each end of the eABF grid range, using a spring constant of $\kappa = 10^5$ kJ/mol/(CV unit)$^2$ to ensure the system explores CV values within a reasonable range.
The settings specific to the CV models are summarized in Table~\ref{tab:cv_param}.
Note that filtering the transition state conformations affects the CV range, as unfiltered transition state conformations would overlap with metastable state conformations.

\begin{table}
\begin{center}
\caption{Enhanced sampling parameters specific to the CV models.}
\label{tab:cv_param}
\begin{tabular}{lcccc}
\toprule
\multicolumn{1}{c}{} & \multicolumn{2}{c}{eABF (DRR)} & \multicolumn{2}{c}{Metadynamics} \\
\cmidrule(lr){2-3}\cmidrule(lr){4-5}
\multicolumn{1}{c}{Model} & {Grid range} & Grid spacing & Grid range & {Gaussian width} \\
\midrule
\textbf{TDA} & $[-8.0, 8.0]$ & 0.10 & $[-8.5, 8.5]$ & 0.20 \\
\textbf{TDA\textsubscript{geo}} (unfiltered) & $[-7.0, 8.0]$ & 0.10 & $[-7.5, 8.5]$ & 0.20 \\
\textbf{TDA\textsubscript{geo}} (filtered) & $[-8.0, 8.0]$ & 0.10 & $[-8.5, 8.5]$ & 0.20 \\
\midrule
\textbf{Reg\textsubscript{geo}} (unfiltered) & $[-0.25, 1.15]$ & 0.01 & $[-0.30, 1.20]$ & 0.02 \\
\textbf{Reg\textsubscript{geo}} (filtered) & $[-0.20, 1.10]$ & 0.01 & $[-0.25, 1.15]$ & 0.02 \\
\bottomrule
\end{tabular}
\end{center}
\end{table}

\subsection{Geodesic interpolation and collective variables training}
\label{sec:appendix_augmentation}

\paragraph{Extraction of folded and unfolded configuration.}
\label{sec:appendix_extraction}
We sampled 5,000 folded and 5,000 unfolded chignolin configurations at 10 ps intervals from their respective 50-ns unbiased simulations.

\paragraph{Geodesic interpolation.}
\label{sec:appendix_geo_interp}
All heavy atoms of chignolin were used for interpolations.
All folded and unfolded state structures were aligned to the first frame of the unfolded state dataset.
For each interpolation calculation, two endpoints are selected from the metastable states and interpolated with a ratio of $t:1-t$, where $t$ is an interpolation parameter sampled from the normal distribution $\mathcal{N}(0.5, 0.05^2)$.
We obtained 5,000 synthetic transition state conformations through geodesic interpolation.

\paragraph{Filtering interpolation dataset.}
We additionally filtered the interpolated conformations using the \textbf{TDA} model (see below) to ensure that the sampled conformations do not overlap with metastable state conformations.
We selected conformations within the \textbf{TDA} CV range of $[-5, 5]$, as in Ray et al.\cite{ray2023deep}, resulting in 2,961 synthetic transition state configurations.

\paragraph{CV model training details.}
\label{sec:appendix_CV_training}
The input to the models consists of the 45 pairwise contacts between the 10 C$_\alpha$ atoms of chignolin. 
To ensure that the CV has continuous derivatives with respect to atomic coordinates, each contact $s_{ij}$ is defined using a switching function given by eq~\eqref{eq:contact}, where $r_{ij}$ denotes the distance between C$_\alpha$ atoms $i$ and $j$, $r_0=0.8$ nm, $n=6$, and $m=12$. 
Consequently, the contact $s_{ij}$ can be interpreted as a continuous version of a coordination number \cite{tribello2014plumed}.
\begin{equation}
s_{i j}=\frac{1-\left(r_{ij}/r_0\right)^n}{1-\left(r_{ij}/r_0\right)^m}
\label{eq:contact}
\end{equation}

We implemented our CVs using the mlcolvar 1.1.0 package \cite{bonati2023unified}, using default settings for model and training unless otherwise specified.
Both the TDA and regression models were implemented as multi-layer perceptrons (MLPs) with layer dimensions of $(45, 24, 12, 1)$ and ReLU activation layers.
The input features were used without additional normalization since the contact features are already normalized to a range of $[0, 1]$.
Following Ray et al.\cite{ray2023deep}, the target distribution parameters were set to $(\bar{\mu}_{\mathcal{U}}, \bar{\mu}_{\mathcal{F}}) = (-7.0, 7.0)$ and $(\bar{\sigma}_{\mathcal{U}}, \bar{\sigma}_{\mathcal{F}}) = (0.2, 0.2)$ for two-state models (\textbf{TDA}), and $(\bar{\mu}_{\mathcal{U}}, \bar{\mu}_{\mathcal{T}}, \bar{\mu}_{\mathcal{F}}) = (-7.0, 0.0, 7.0)$ and $(\bar{\sigma}_{\mathcal{U}}, \bar{\sigma}_{\mathcal{T}}, \bar{\sigma}_{\mathcal{F}}) = (0.2, 1.5, 0.2)$ for three-state models (\textbf{TDA\textsubscript{geo}}).
Subscripts $\mathcal{U}, \mathcal{T}, \mathcal{F}$ denote unfolded, transition, and folded states, respectively.

We split the dataset into 80\% for training and 20\% for validation.
The models were trained for 2,000 epochs using the Adam optimizer \cite{kingma2017adam} with a learning rate of 0.001, and we applied early stopping with a patience of 50 epochs based on the validation loss.

\subsection{Structural analysis}
\label{sec:appendix_struc}
\begin{figure*}[!ht]
    \includegraphics[width=\textwidth]{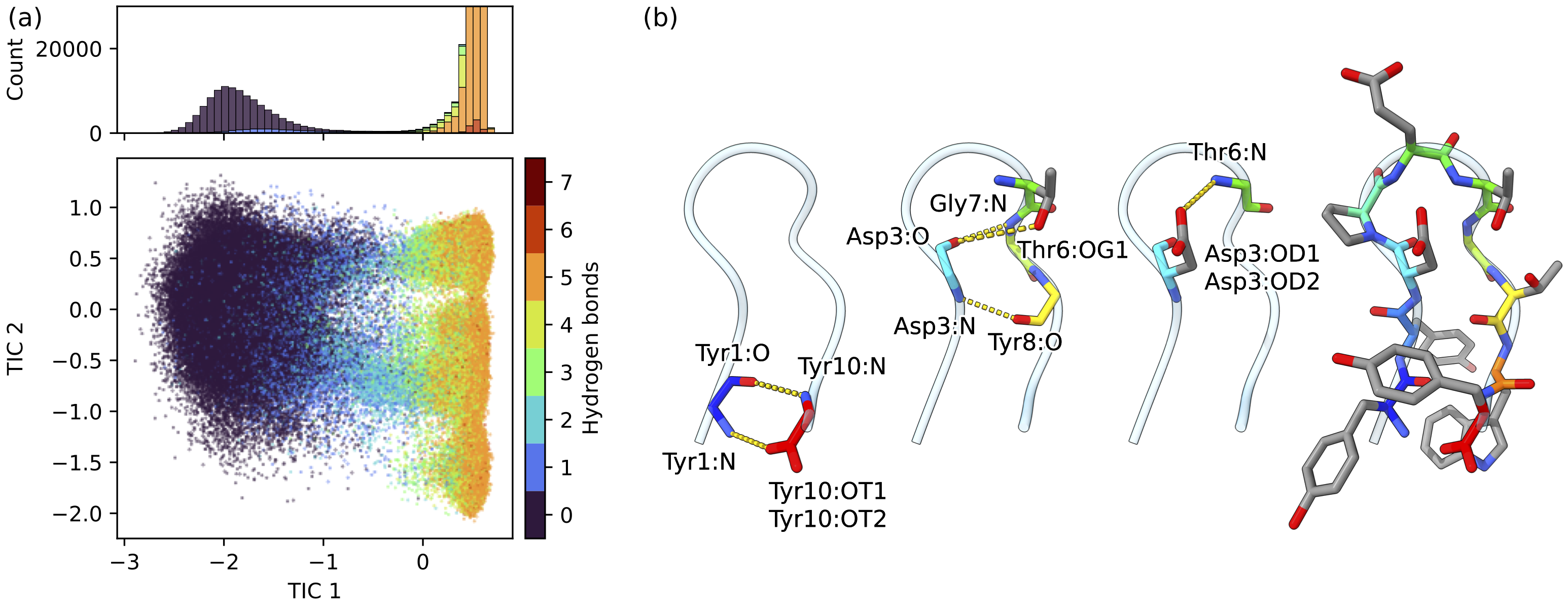}

    \caption{\textbf{(a)} Projection of conformations from the unbiased reference trajectory onto the first two time-lagged independent components (TIC~1 and~2), with the distribution along TIC~1, colored by the number of hydrogen bonds in the folded state.
    \textbf{(b)} Definition of the eight hydrogen bonds observed in the folded state of chignolin.
    The three structures to the left illustrate the donor and acceptor for each hydrogen bond, while the rightmost structure provides the complete structure of chignolin for reference.
    For the cases where hydrogen bonds involve two equivalent oxygens in carboxylates (Asp3:OD1/OD2 and Tyr10:OT1/OT2), a single instance is depicted in the structural diagrams.}
    \label{fig:TICA_hydrogen_bond}
\end{figure*}

\paragraph{Folded state hydrogen bonds.}
In addition to time-lagged independent component analysis, we devised a structure-based method to classify conformations into folded, transition, and unfolded states.
We collected eight distinctive hydrogen bonds observed in the folded state, Tyr1:N--Tyr10:OT1, Tyr1:N--Tyr10:OT2, Asp3:N--Tyr8:O, Thr6:OG1--Asp3:O, Thr6:N--Asp3:OD1, Thr6:N--Asp3:OD2, Gly7:N--Asp3:O, and Tyr10:N--Tyr1:O, as depicted in Figure~\ref{fig:TICA_hydrogen_bond}(b).
The hydrogen bonds are detected in the conformation based on the two criteria: (1) acceptor and donor atom distance within 3.5~\AA{} and (2) acceptor--hydrogen--donor angle larger than 110$^\circ$.
The formation of chignolin $\beta$-hairpin involves two key structural aspects: the pairing of beta sheets at the terminal regions and the turn formation \cite{mckiernan2017modeling,sobieraj2022granger}.
The transition from the unfolded to the folded state consists of the simultaneous formation of these two structural features, suggesting a classification threshold of two hydrogen bonds.
The projection of conformations from the reference trajectory onto the slowest modes (Figure~\ref{fig:TICA_hydrogen_bond}(a)) validates that the two metastable states can be distinguished based on the number of hydrogen bonds, with the transition region along TIC~1 aligning with the two hydrogen bonds.

\section{Additional simulation analysis}

\subsection{Time evolution of the CV}

\begin{figure*}[!ht]
    \includegraphics[width=\textwidth]{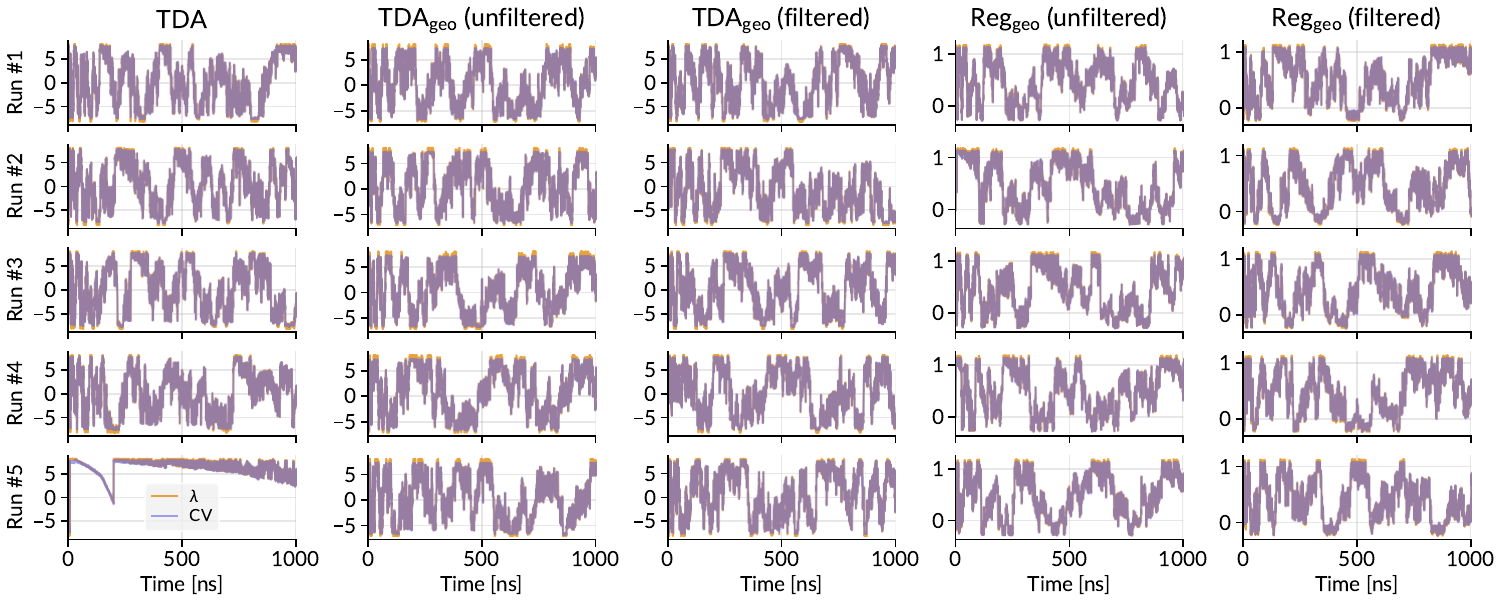}
    \caption{Time evolution of the CV and the fictitious particle position $\lambda$ for WTM-eABF enhanced sampling simulations using different CV models.
    Each row represents one of the five distinct initial configurations described in Section~\ref{sec:simulation_details}.}
    \label{fig:time_cv}
\end{figure*}

\subsection{C$_\alpha$--TICA projections of the simulated trajectories}

\label{sec:appendix_tica}
\begin{figure*}[!ht]
    \includegraphics[width=\textwidth]{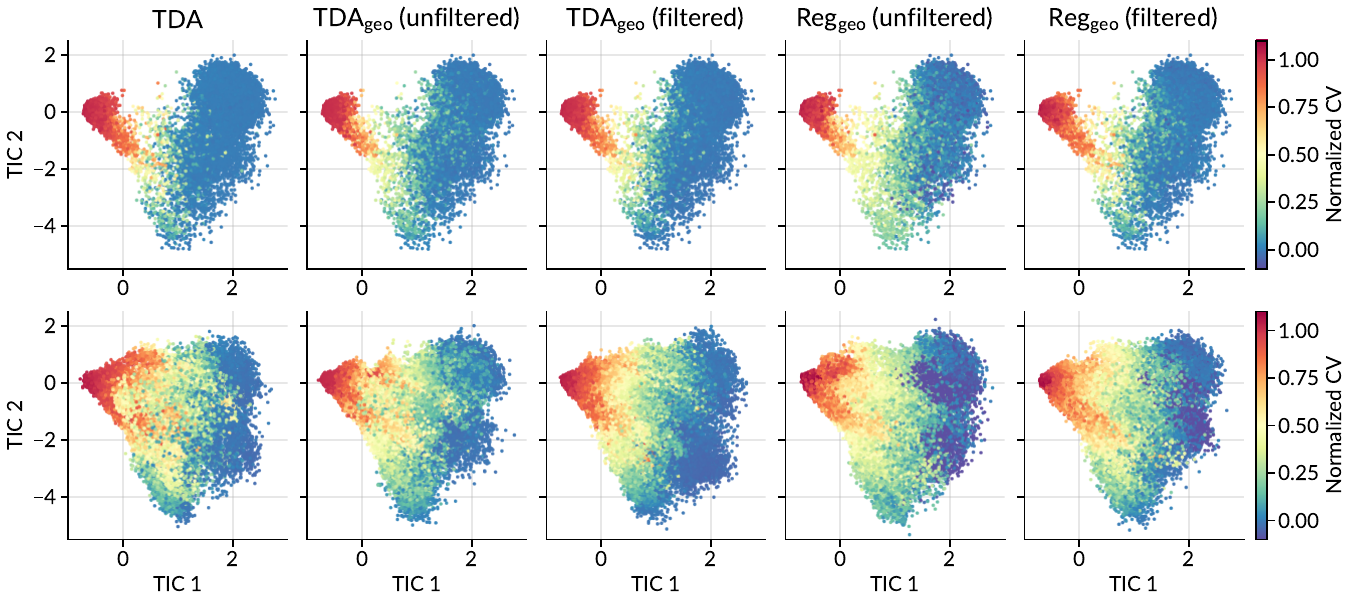}
    \caption{The TICA analysis was performed on the C$_\alpha$--C$_\alpha$ distances from the unbiased trajectory as inputs. Projections of the conformations from the unbiased reference trajectory (upper rows) and WTM-eABF enhanced sampling trajectories using the different ML-CVs onto the first two time-lagged independent components, colored based on the normalized CV value. The CV values were scaled such that the PMF minima of the unfolded and folded basins correspond to 0 and 1, respectively.}
    \label{fig:tica_cv_bb}
\end{figure*}

In the main text, Figure~5 displays the TICA projections colored based on the normalized CV value, where TICA was conducted using all heavy-atom 3-D coordinates as input features.
Additionally, Figure~\ref{fig:tica_cv_bb} shows CV-colored TICA projections generated only from C$_\alpha$--C$_\alpha$ distances. 
In the C$_\alpha$--TICA plot, the upper-left region corresponds to the folded state, while the right region corresponds to the unfolded state. 
Both visualizations yield a consistent conclusion: the \textbf{TDA} models exhibit misclassification behavior, particularly by inaccurately predicting the transition state as a folded state.

\section{Computational cost}

\label{sec:computational_cost}
\begin{figure*}[!ht]
    \includegraphics[width=0.4\textwidth]{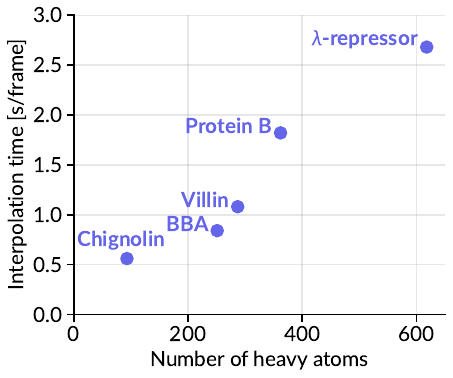}
    \caption{Wall clock time per interpolated frame for geodesic interpolation on a single NVIDIA Volta V100 32GB GPU for five different proteins, plotted against the number of heavy atoms.}
    \label{fig:interp_time}
\end{figure*}

Here, we report the computational cost of geodesic interpolation using a single NVIDIA Volta V100 32GB GPU paired with 20 cores of an Intel Xeon Gold 6248 CPU.
In addition to chignolin (CLN025, PDB 5AWL) studied in the main text, we benchmarked four larger proteins: BBA (1FME), Villin (2F4K), Protein B (1PRB), and $\lambda$-repressor (1LMB), to evaluate the scalability of the interpolation method.
These proteins have 10, 28, 35, 47, and 80 residues, respectively.

As detailed in Section \ref{sec:appendix_augmentation}, we performed geodesic interpolation calculations for each protein and reported the wall clock time per interpolated frame against the number of heavy atoms in Figure~\ref{fig:interp_time}.
For chignolin, interpolating 5,000 frames took approximately 40 minutes, which is about half the time of the 50 ns unbiased simulation (80 minutes) needed to extract 5,000 metastable state frames in our settings.
Additionally, the computational cost scales almost linearly with the number of heavy atoms, suggesting that this method is computationally feasible for larger systems.

\bibliography{main}